# Photonic Generation of High Power, Ultrastable Microwave Signals by Vernier Effect in a Femtosecond Laser Frequency Comb

Khaldoun Saleh*, Jacques Millo, Baptiste Marechal, Benoît Dubois, Ahmed Bakir, Alexandre Didier, Clément Lacroûte & Yann Kersalé

FEMTO-ST, Univ. Bourgogne Franche-Comté, CNRS, ENSMM, 26 Rue de l'Épitaphe, 25030 Besançon cedex, France.
*khaldoun.saleh@femto-st.fr

**ABSTRACT**

Optical frequency division of an ultrastable laser to the microwave frequency range by an optical frequency comb has allowed the generation of microwave signals with unprecedently high spectral purity and stability. However, the generated microwave signal will suffer from a very low power level if no external optical frequency comb repetition rate multiplication device is used. This paper reports theoretical and experimental studies on the beneficial use of the Vernier effect together with the spectral selective filtering in a double directional coupler add-drop optical fibre ring resonator to increase the comb repetition rate and generate high power microwaves. The studies are focused on two selective filtering aspects: the high rejection of undesirable optical modes of the frequency comb and the transmission of the desirable modes with the lowest possible loss. Moreover, the conservation of the frequency comb stability and linewidth at the resonator output is particularly considered. Accordingly, a fibre ring resonator is designed, fabricated, and characterized, and a technique to stabilize the resonator's resonance comb is proposed. A significant power gain is achieved for the photonically generated beat note at 10 GHz. Routes to highly improve the performances of such proof-of-concept device are also discussed.

## Introduction

For more than twenty years, optics, nonlinear optics and microwave-photonics have proven themselves as elegant and reliable solutions for the development of microwave and millimetre-wave sources featuring very high stability and spectral purity[1-6]. The advent of optical frequency combs generated by femtosecond lasers in the late 1990s[7-9] has particularly revolutionized the time and frequency metrology domain. These femtosecond lasers have allowed the establishment of "bridges" linking distant parts in the electromagnetic spectrum (gamma ray, X-ray, UV, IR, microwave, …). Following this discovery, the principle of optical frequency division of an ultrastable laser to much lower frequencies (microwave frequency range) by an optical frequency comb has been demonstrated[10]. Through this process, photonic generation of microwave signals with unprecedently high spectral purity and stability is achieved. Such performances become increasingly crucial in several areas of applications, particularly in communication systems, radars, signal processing, radio astronomy, satellites, GPS navigation, spectroscopy and in time and frequency metrology.

In this approach, the frequency of a continuous wave laser is first stabilized onto an optical mode of an ultrastable optical cavity. This ultrastable laser is then used to stabilize the repetition rate frequency ($f_{rep}$) of an optical frequency comb (OFC) generated by a self-referenced femtosecond laser. This transfers the spectral purity and the stability of the ultrastable laser to the frequency comb modes. The photodetection of this OFC provides a RF frequency comb at $f_{rep}$ and its harmonics. A bandpass microwave filter then selects any desired microwave frequency and rejects the other components of the RF frequency comb. Here, the OFC acts as an optical-to-microwave stability bridge, since the stability of the ultrastable laser is (ideally) conserved in fractional values in the microwave domain. It also acts as an optical-to-microwave phase noise divider, as the generated microwaves benefit from an extremely low phase noise level, corresponding theoretically to a reduction by $20\log R$ in the ultrastable laser phase noise, where $R$ is the frequency ratio between the ultrastable laser and the generated microwave harmonic ($20\log R$ = 86 dB between a laser at 200 THz and a microwave harmonic at 10 GHz). Outstanding results are obtained today with this process in terms of stability and absolute phase noise level[11, 12, 13] (fractional frequency stability below 6.5x10$^{-16}$ at one second and phase noise level below -105 dBc/Hz and -167 dBc/Hz, respectively, at 1 Hz and 10 kHz offset frequencies from a 12 GHz carrier[14]). Unfortunately, if no external OFC repetition rate multiplication device is used, this 10 GHz photonically generated microwave signal will suffer from a very low power level (around -30 dBm with standard PIN photodiodes). This is because the low mode spacing of the OFC ($f_{rep} \approx$ hundreds of MHz) leads to a rapid saturation of the photodiode used for the conversion of the OFC into the microwave frequency domain.

In microwave photonic generators, the direct generation of a high-power microwave signal without microwave amplification will overcome the need for microwave amplifiers in the generator scheme. In this case, the signal-to-noise ratio is improved, the phase noise floor is reduced and the noise contribution of the microwave amplifiers to the generated signal is avoided. High-power handling and high-linearity photodiodes are being developed nowadays[15], which help increase the power of the photonically generated microwaves, yet at a very high cost. Regarding the $f_{rep}$ multiplication devices, in 1989, T. Sizer[16] has demonstrated the beneficial use of the optical resonance comb (ORC) of a Fabry-Pérot optical cavity to filter an OFC selectively and repetitively, therefore increasing its $f_{rep}$ up to 12 times. Later, the idea has been particularly studied by researchers in the field of astronomical spectroscopy for spectrographic calibration[17, 18], when high stability OFCs became available. Therefore, if $f_{rep}$ is multiplied up to several tens of GHz, while maintaining a large spectrum of the OFC, the multiplied OFC turns into a perfect light-frequency ruler, ideal for spectrographs calibration. In photonic microwave generators, compared to the case of an unfiltered OFC, up to 11 dB power gain at the 10 GHz generated microwave harmonic has been obtained by using a Fabry-Pérot filtering cavity with 5 GHz free spectral range ($FSR$) to filter the OFC[19]. Nevertheless, the Fabry-Pérot filtering cavity approach suffers from a limited output comb linewidth due to the cavity chromatic dispersion. This affects the pulse train characteristics and is hard to be managed in such cavities. Moreover, this approach suffers from a reduction in the average output optical power by at least the same $f_{rep}$ multiplication factor $M$ (selection of every $M$ th mode of the femtosecond frequency comb). Complex cascaded Fabry-Pérot cavities architecture can be however implemented to recycle this lost optical power[16, 18].

Another elegant low loss solution to increase the femtosecond laser repetition rate is the use of cascaded fibred Mach-Zehnder interferometers (MZIs) for pulses temporal interleaving[20]. Compared to an unmultiplied pulse train, up to 18 dB power gain at 12 GHz microwave harmonic has been obtained by cascading 3 MZIs and a high linearity photodiode. Nevertheless, very high precision is needed in fabricating such fibred MZIs (less than few hundreds of micrometres margin of error), especially if they are intended to be used in a low noise photonic microwave generator[21]. Integrated free-space MZIs could be however a way to circumvent this problem.

Like the Fabry-Pérot cavity, however with numerous practical advantages, the single transverse ORC of the ring resonator can be more easily engineered to filter an OFC selectively and repetitively. Direct coupling add-only ring resonator's ORC modes have been previously used as notch repetitive filters to double the repletion rate of an OFC[22]. In this paper, we present theoretical and experimental investigations performed on the beneficial use of a double directional coupler add-drop fibre ring resonator (FRR) and Vernier effect to filter an OFC of a fibred femtosecond laser selectively and repetitively and increase its 250 MHz repetition rate up to 10 GHz. By means of a computer-aided design software and a FRR model, a thorough study is conducted to identify the Vernier solutions allowing better $f_{rep}$ multiplication efficiency. Following, a dispersion managed FRR increasing $f_{rep}$ up to 10 GHz is designed, fabricated, and tested, and a full femtosecond comb linewidth conservation is obtained. Furthermore, compared to an unfiltered OFC, up to 12 dB power gain at 10 GHz microwave harmonic is achieved. Routes to highly improve the performances of such proof-of-concept device are also proposed and discussed.

## Results
### 10 GHz Vernier Effect with a Fibre Ring Resonator
By Vernier effect, the transmission of a coupled system occurs when the modes of the subsystems coincide. As illustrated in Figure 1, the combination of the OFC of the femtosecond laser and the ORC of the FRR (see Methods section) selectively preserves or suppresses (completely or partially) optical modes of the OFC. This increases the OFC's repetition rate to a given frequency of interest equal to the coupled system's total free spectral range:

$$FSR_{tot} = N_1 \times f_{rep} = N_2 \times FSR_{FRR} \quad (1)$$

where, $N_1$ and $N_2$ are integers, and $FSR_{FRR}$ is the FRR's free spectral range.

In this paper, the frequency of interest we are seeking is 10 GHz, but much higher frequencies are also within reach. To be able to use the ORC of the FRR as a selective and repetitive filter to respectively preserve and reject desirable and undesirable modes of the OFC, the $FSR_{FRR}$ must be precisely calculated to get the desired Vernier effect. Following the abovementioned relation and by fixing $FSR_{tot}$ to 10 GHz ($f_{rep}$ and $N_1$ respectively equal to 250 MHz and 40), several solutions can be calculated for the $FSR_{FRR}$ for different values of $N_2$ (1 to $\infty$). As we will see later, not all the solutions provide the same preservation and rejection ratio of desirable and undesirable modes of the OFC. Moreover, even if the solution obtained with $N_2 = 1$ and thus $FSR_{FRR}$ = 10 GHz appears to be the simplest one, the FRR will be technically very difficult to fabricate in this case.

In the following, we qualify $\beta = \sum P_{Desirable} / \sum P_{Undesirable}$ as the optical power ratio between the OFC's desirable modes, repetitively spaced by 10 GHz, and the undesirable modes in-between. Without selective filtering by a FRR, $\beta$ = -12.9 dB for an OFC with 250 MHz repetition rate. In our approach, the aim is to obtain a high $\beta$ value using a low insertion loss FRR, therefore, well preserved OFC desirable modes and highly rejected undesirable modes.

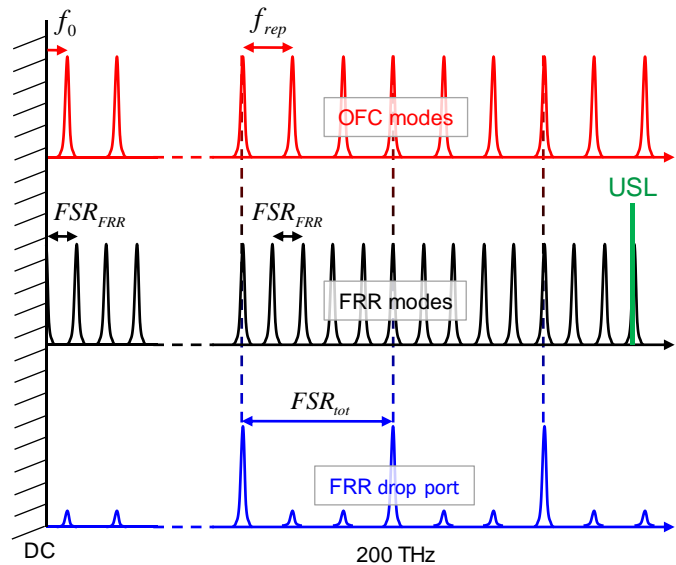

**Figure 1. Vernier effect obtained by combining the OFC of the femtosecond laser with the ORC of the FRR to selectively preserve or suppress optical modes of the OFC and increase its repetition rate up to a given frequency of interest.** The ultrastable laser (USL) is used to lock the ORC.

A Vernier solution consisting in multiplying the repetition rate of the OFC up to 10 GHz was studied using a computer-aided design software (Agilent ADS) and a FRR model[23], greatly simplifying the simulation of complex FRR based architectures. Since the ORC of the FRR is used as a selective and repetitive filter, our studies were focused on two of its characteristics: the high out-of-band rejection and the low insertion loss. Both parameters are directly related to the overall FRR losses and to the couplers coupling coefficients (see Methods section). In addition, when propagating through the FRR, the optical pulse duration increases due to chromatic dispersion. The FRR dispersion must be therefore managed and compensated to reduce the pulse duration down to its original width.

In the following, studies were performed on a FRR design including a dispersion compensation fibre section to prevent any degradation in the OFC spectrum due to chromatic dispersion in the FRR. We have simulated the Vernier solutions for different values of $N_2$, while setting 0.3 dB couplers excess loss, optimized 10% symmetrical coupling coefficient[24], 0.02 dB splices loss, 0.5 dB dispersion compensation fibre section loss and 0.2 dB/km fibre attenuation. The simulations helped us to identify different Vernier patterns for each Vernier solution ($FSR_{tot}$ = {1, 1.25, 2, 2.5, 5, 10 GHz}).

The optical power ratio $\beta$ has been calculated for all the Vernier solutions over a 10 GHz optical bandwidth as it reflects the power ratio over the whole OFC spectrum. Interestingly, it has been found that each Vernier solution has a constant $\beta$, regardless the formed Vernier pattern. For example, different 2 GHz Vernier patterns are obtained for $N_2$ = {25, 35, 45, 55, 65…} but with a constant $\beta$ = -3.8 dB. Moreover, it has been found that the higher the Vernier $FSR_{tot}$ is, the higher is $\beta$. As we can see from Figure 2 (a), $\beta$ = -6.6 dB for a 1 GHz Vernier solution, compared to a $\beta$ = -1.2 dB for a 10 GHz Vernier solution.

The most evident patterns for the abovementioned Vernier solutions were simulated over a 10 GHz optical bandwidth, while considering an input OFC having a 0 dBm optical power per mode, and are given in Figure 2 (b).

It is noteworthy that when the FRR length increases, its full width at half maximum ($FWHM$) decreases and therefore the $Q$ factor increases. Accordingly, the FRR equivalent length increases[25]. On the other hand, the $FSR$ of the FRR will decrease. Although it can be considered intuitive, the behaviour of these parameters should be considered when designing the FRR for Vernier effect and OFC filtering.

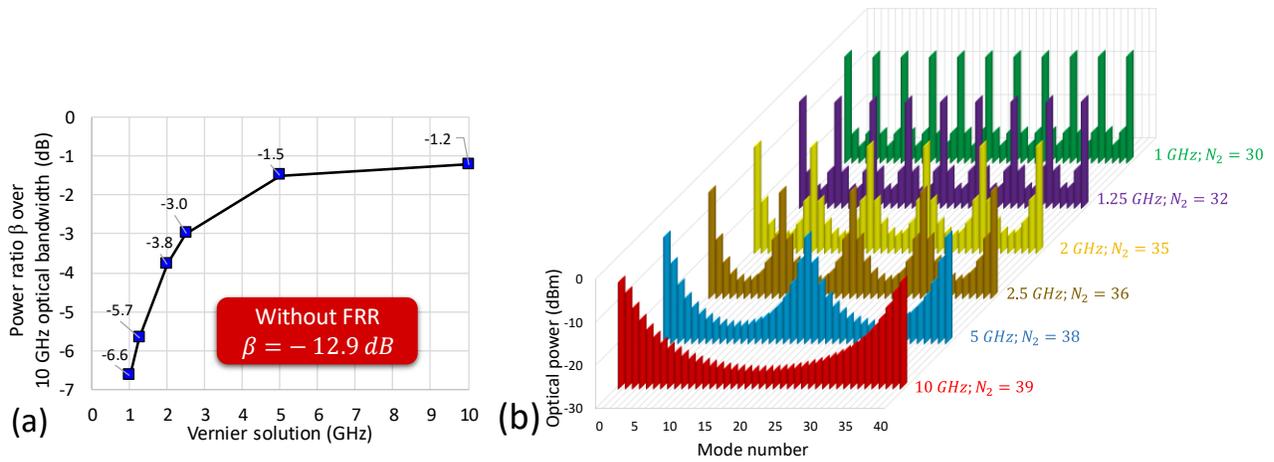

**Figure 2. Vernier solutions for different values of** $N_2$. (a) Optical power ratio $\beta$ between the OFC desirable modes, spaced by 10 GHz, and the undesirable modes in-between, calculated for different Vernier solutions over a 10 GHz optical bandwidth. (b) Most evident patterns for different Vernier solutions simulated over a 10 GHz optical bandwidth, while considering an input OFC having a 0 dBm optical power per mode. The mode number scale refers to the OFC modes included in the 10 GHz optical bandwidth, where modes 0 and 40 are the desirable optical modes spaced by 10 GHz and the modes from 1 to 39 are the undesirable optical modes in-between.

Since the 10 GHz Vernier solution has been theoretically found to have the highest power ratio among the other solutions, ($\beta$ = -1.2 dB), and thus an optical power ratio gain ~ 11 dB compared to an unfiltered OFC ($\beta$ = -12.9 dB), we have searched for the best 10 GHz Vernier pattern that simplifies the FRR fabrication and use. Figure 3 (a) shows some of the 10 GHz Vernier solution patterns. Here, a symmetry in the patterns repeatability is noticeable, whether the $N_2$ value is higher or lower than $N_1$ value ($N_1$ = 40) by the same integer number (e.g. $N_2$ =39 and $N_2$ = 41, $N_2$ = 37 and $N_2$ = 43, etc.).

As already mentioned these different Vernier patterns provide the same $\beta$. However, they do not have the same difficulties related to FRR fabrication and use. Since the FRR dispersion must be managed to get a transparent FRR in terms of chromatic dispersion regarding the OFC, the FRR length should be long enough to allow the insertion of a dispersion compensation fibre section. On the other hand, increasing the FRR length will increase the FRR's $Q$, consequently increasing the FRR susceptibility to external perturbations and therefore the need for a highly performant locking scheme to ensure the ORC stability. Moreover, an increase in the $Q$ factor, and consequently in the FRR equivalent length, will increase the risk of generation of nonlinear optical effects inside the FRR (stimulated Brillouin scattering, Four wave mixing, etc…[26]). As a result, a trade-off must be found, bearing in mind the necessity of OFC spectrum and linewidth conservation after being filtered by the FRR, and a linewidth conservation control using the available low resolution optical spectrum analyser (OSA; 2.5 GHz resolution). Therefore, a Vernier pattern visible with such OSA must be chosen.

From Figure 3 (b), we can see that the Vernier pattern obtained with $N_2$ = 67 should be visible using a low resolution OSA and should therefore allow an accurate check on the OFC spectrum and linewidth conservation, and on a possible ORC walk-off phenomenon during the selective filtering process. Moreover, the FRR length (~1.36 m) is adequate to allow the insertion of a dispersion compensation fibre section in the FRR during its fabrication. The $FSR_{FRR}$ will be equal to ~149 MHz in this case.

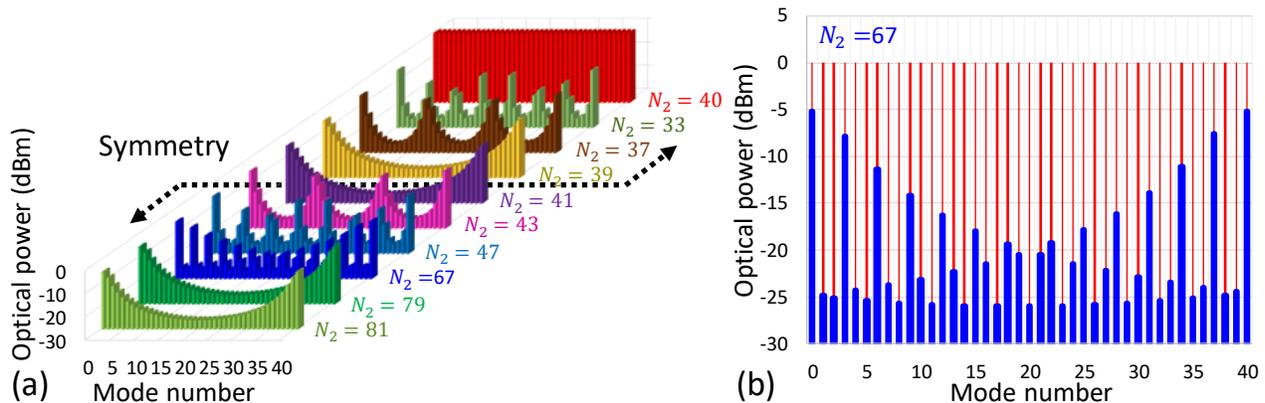

**Figure 3. Simulated Vernier patterns obtained for the 10 GHz Vernier solution, while considering an input OFC having a 0 dBm optical power per mode**. (a) Some of the patterns of the 10 GHz Vernier solution, where a symmetry in the patterns repeatability is noticeable, whether the $N_2$ value is higher or lower than $N_1$ value (40) by the same integer number (the solution where $N_2$ = 40 is exposed in the back for more visibility of the symmetry in the patterns repeatability). (b) The Vernier pattern obtained when $N_2$ = 67, compared to the case where no FRR is used.

**Experimental characterization**
Based on the above theoretical results, a dispersion compensated FRR with a total length ensuring a $FSR_{FRR}$ = 10 GHz/67 has been designed and fabricated (see Methods section). A 5.2 dB insertion loss and 20.7 dB out-of-band rejection were measured and are in excellent agreement with our simulation results.

The experimental setup that we used to characterize the OFC's $f_{rep}$ multiplication by the compensated FRR is depicted in Figure 4. In this setup, a femtosecond laser OFC is first self-referenced by extracting its carrier envelope offset's signal (CEO; $f_0$; see Figure 1) and stabilizing it to a 10 MHz hydrogen maser signal (see Figure 4). The self-referenced femtosecond laser output is sent to the first input of the compensated FRR (20 dBm average input optical power). This is done through a first dispersion compensation fibre section (ensuring a dispersion compensation between the femtosecond laser and the compensated FRR), a variable optical attenuator, a polarization controller (ensuring an efficient excitation of the TE or TM ORC modes in the compensated FRR) and an optical circulator (ensuring a good redirection and isolation of the counterpropagating femtosecond laser and ultrastable laser signals). The filtered OFC at the first output of the compensated FRR is then sent through a second dispersion compensation fibre section either to an OSA, or to be photoconverted by a fast photodiode and then sent to an ultrafast oscilloscope or to an electrical spectrum analyser (ESA). The average optical power at the compensated FRR first output is monitored by an optical power meter. The other part of the OFC at the second input of the compensated FRR is sent through an optical circulator to an OFC's repetition rate stabilization unit.

On the other hand, the signal of an ultrastable laser, stabilized to a high finesse Fabry-Pérot cavity, is sent simultaneously to the OFC's $f_{rep}$ stabilization unit and to the second input of the compensated FRR through a polarization controller and an optical circulator. The ultrastable laser part at the first input of the FRR is deviated from the incident femtosecond laser by the optical circulator. The other ultrastable laser part at the second output of the compensated FRR is detected by a slow photodiode. The detected signal serves as an error signal for the ORC locking using the side-of-fringe locking technique (see

Methods section). To take full benefit of the Vernier effect, it is noteworthy that the resonance comb of the compensated FRR is first locked to the ultrastable laser. Afterwards, $f_0$ and $f_{rep}$ are varied to optimize the Vernier effect at the compensated FRR first output and then locked. The polarization of the input OFC is of course optimized and later fixed to take full benefit of the selective filtering by the ORC. The overall setup remains locked for several months without human intervention, while preserving all the characteristics of the output optical and microwave signals (multiplied $f_{rep}$, average optical power, microwave harmonics power levels).

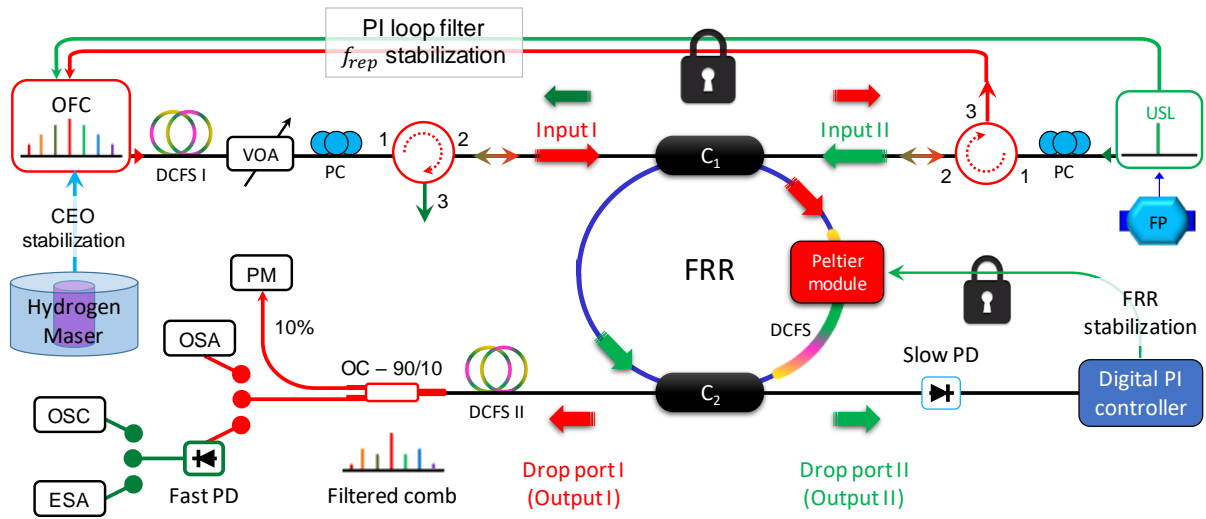

**Figure 4. Experimental setup used to characterize the OFC's $f_{rep}$ multiplication by Vernier effect using the FRR.** OFC: femtosecond laser optical frequency comb; DCFS: dispersion compensation fibre section; VOA: variable optical attenuator; PC: polarization controller; FP: Fabry-Pérot cavity; USL: ultrastable laser; OC: optical fibre coupler; PM: power meter; OSA: optical spectrum analyser; PD: photodiode; OSC: ultrafast oscilloscope; ESA: electrical spectrum analyser; PI: digital proportional–integral controller.

An average optical power loss of almost 15.5 dB was measured at the compensated FRR output. This power loss is due to the 5.2 dB insertion loss of the compensated FRR and to the filtering of undesirable modes. Here, it should be noted that the recycling of the lost optical power due to the filtering of undesirable modes, by cascading several compensated FRRs, could be less complex than in the case of Fabry-Pérot cavities[16].

Afterwards, the OFC spectrum has been recorded, at 3 dBm incident average optical power, with and without compensated FRR filtering using the OSA (RBW= 2.5 GHz). This is to ensure that a full transparency of the compensated FRR has been achieved in terms of chromatic dispersion regarding the OFC. The results are depicted in Figure 5 (a). The wide span (18.7 THz; 150 nm) optical spectra prove a total conservation of the OFC bandwidth at the compensated FRR output. A zoom into the spectrum shows the formation of the 10 GHz Vernier pattern [Figure 5 (b)]. This experimental result is in very good agreement with our simulation results. In addition, this pattern was continuously repetitive over the whole optical spectrum without deformation and no ORC walk-off phenomenon was observed over the entire bandwidth. This proves the validity of our FRR dispersion compensation approach.

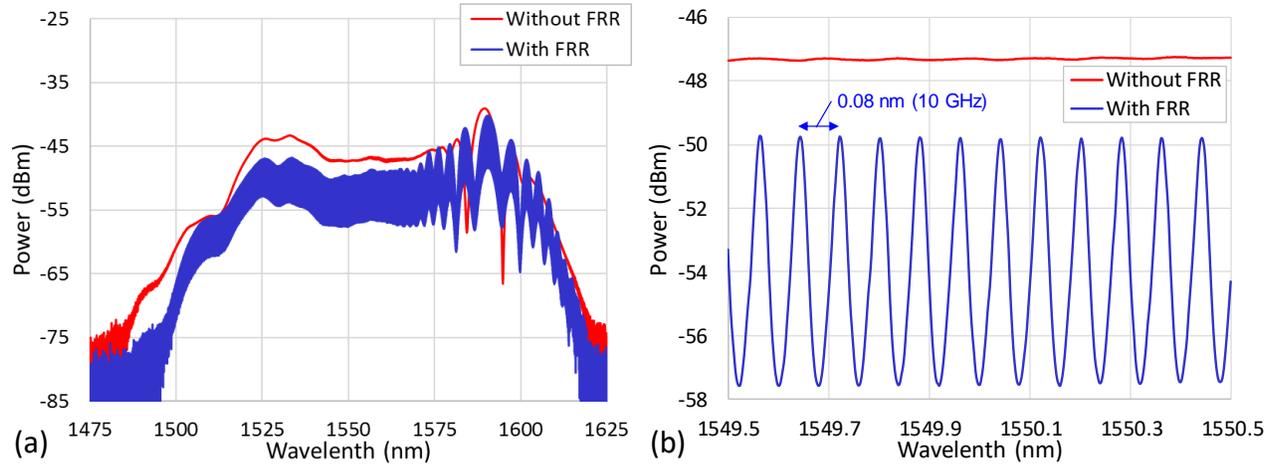

**Figure 5. Optical domain characterization.** (a) Optical spectra recorded with and without the compensated FRR, at 3 dBm incident average optical power, confirming a total conservation of the OFC linewidth at the compensated FRR output. (b) Zoom into the recorded optical spectra showing the 10 GHz Vernier pattern formation. OSA: RBW = 2.5 GHz.

The resulting microwave frequency comb was then recorded with the ESA (RBW=100 kHz), at 3 dBm incident average optical power. The characteristic selective repetitive filtering of the chosen 10 GHz Vernier pattern is clearly visible when using the compensated FRR and is in very good agreement with our simulation results [see Figure 6 (a)(b)]. Moreover, the 9.9978 GHz generated frequency means that a less than +1 mm fabrication error has been achieved without sophisticated length measurement and fabrication instruments. This was achieved even though 10 cleaving and 5 splices have been performed (see Methods section) for the five fibre sections constituting the compensated FRR. Here, it is also noteworthy to mention that, as for the 10 GHz harmonic, a high power level is also obtained for the 20 GHz microwave harmonic. This proves the possible use of our $f_{rep}$ multiplication approach to photonically generate high power and high stability millimetre waves.

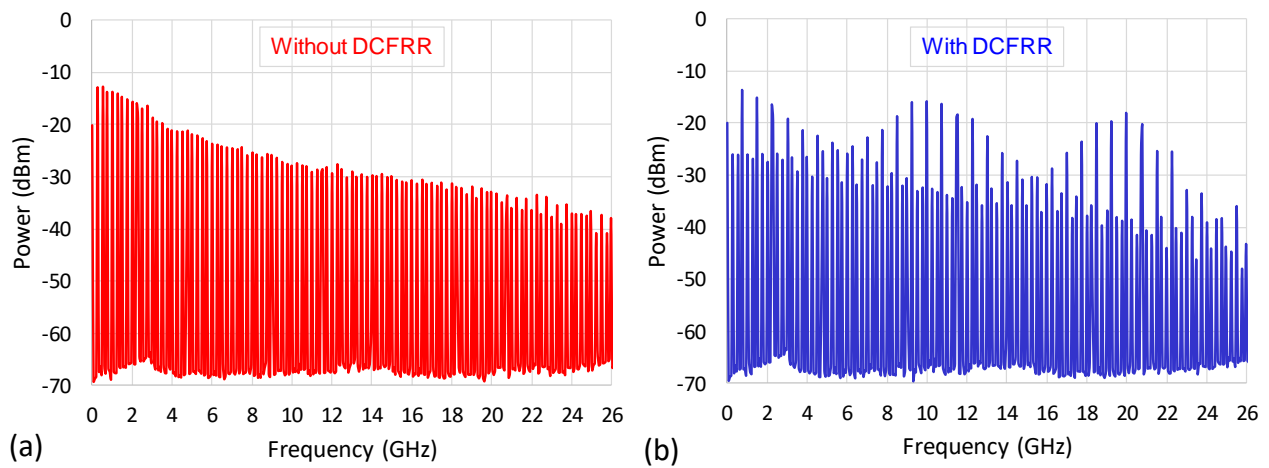

**Figure 6. Microwave domain characterization.** Microwave spectra recorded (a) without and (b) with the compensated FRR, at 3 dBm incident average optical power, confirming the chosen 10 GHz Vernier pattern formation. ESA: RBW = 100 kHz and VBW = 100 kHz. DCFRR: dispersion compensated FRR.

The photodiode output has been also recorded with the ultrafast oscilloscope at the same incident average optical power level. The results depicted in Figure 7 (a) show mainly a division of the pulse train period by 3 (1.33 ns). This is primarily due to the abundant presence of the 750 MHz frequency and its harmonics in the Vernier pattern we have chosen, besides the 10 GHz frequency and its harmonics. A zoom into the pulses at the compensated FRR output reveals the presence of this 10

GHz frequency and its harmonics (0.1 ns period) [see Figure 7 (b)]. Aside from that, the variation in the pulses intensity at the FRR output is due to the ringing effect, same as the effect observed in Fabry-Pérot cavities[16] and known as "rattling plate" effect. As previously suggested[16], this effect can be reduced by increasing the FRR finesse.

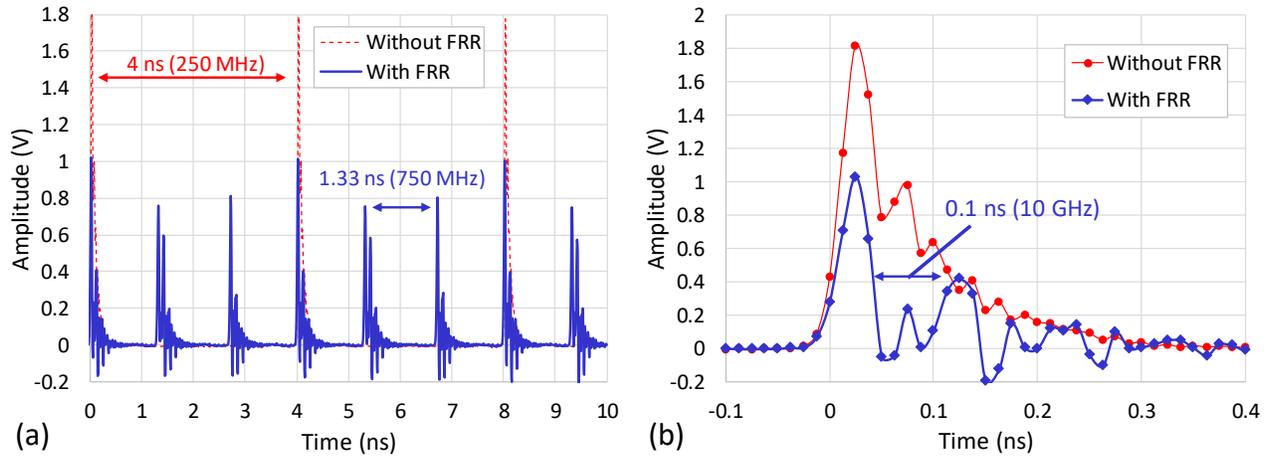

**Figure 7. Time domain characterization.** (a) Electrical waveforms recorded with the ultrafast oscilloscope, with and without the compensated FRR, at 3 dBm incident average optical power. (b) A zoom into the pulses at the compensated FRR output reveals the presence of the 10 GHz frequency and its harmonics (0.1 ns period). Ultrafast oscilloscope sampling rate = 80GS/s.

The microwave power has been measured for the 40[th] harmonic at 10 GHz of the OFC versus the incident average optical power on the fast photodiode and the results are depicted in Figure 8. These results demonstrate the clear improvement in the photodiode linearity thanks to the OFC filtering by the compensated FRR. A microwave power gain up to 12 dB is obtained and the photodiode deviation from linearity at high average optical power levels is clearly reduced.

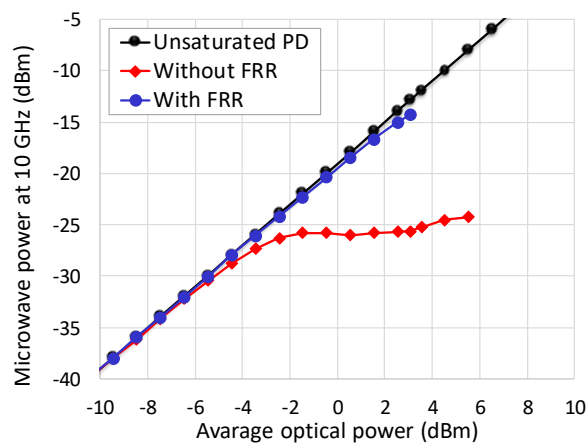

**Figure 8. Microwave power characterization.** Microwave power measurements performed at 10 GHz versus the incident average optical power, obtained with and without the compensated FRR. The case of an ideal photodetection is added for comparison.

## Discussion

As we can see from the above results, the compensated FRR provided very good performances in terms of microwave power gain, photodiode linearity improvement and OFC spectrum and linewidth conservation. Moreover, with 20 dBm average input optical power, no sign of any additional nonlinear losses and no sign of any nonlinear phenomena were observed. This proves that the compensated FRR is an efficient "single" device for OFC repetition rate multiplication. In addition, it is noteworthy that even better results could be further obtained using such a device.

For instance, the compensated FRR intrinsic loss, affecting the resonator's characteristics, is mainly due to the dispersion compensation fibre section loss (~0.5 dB) and it can be highly reduced. This can be done either by using the splice tapering method[27] to further improve the splices quality between the dissimilar fibres composing the dispersion compensation fibre section or by completely removing the dispersion compensation fibre section and using a zero-dispersion fibre to fabricate the entire FRR (which also abridges the FRR design). Both solutions should highly improve the FRR characteristics (lower insertion loss and higher out-of-band rejection) and therefore the repetition rate multiplication process using the Vernier effect.

We have theoretically compared both the compensated FRR and the zero-dispersion FRR performances using the ADS software and the FRR model. The Vernier patterns obtained for $N_2$ = 39 for both cases, compared to the case where no FRR is used, clearly show the important enhancement that one can expect if a zero-dispersion FRR is used as an OFC repetition rate multiplier instead of the compensated FRR (see Figure 9). Here, less insertion loss is obtained for the desirable modes (< 3dB) while much higher rejection is obtained for the undesirable modes (up to 30 dB). Moreover, a 20 dB optical power ratio gain has been calculated for the zero-dispersion FRR compared to the case where no FRR is used (~11 dB optical power ratio gain in the case of the compensated FRR).

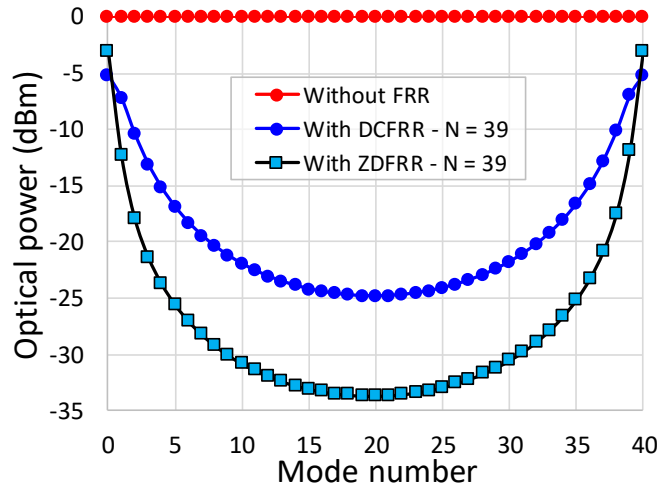

**Figure 9. Simulated Vernier patterns obtained when $N_2$ = 39 with a compensated FRR and a zero-dispersion FRR, compared to the case where no FRR is used, considering an input OFC having a 0 dBm optical power per mode**. The mode number scale refers to the OFC modes included in the 10 GHz optical bandwidth, where modes 0 and 40 are the desirable optical modes spaced by 10 GHz and the modes from 1 to 39 are the undesirable optical modes in-between. DCFRR: dispersion compensated FRR; ZDFRR: zero-dispersion FRR.

Furthermore, for the photonic generation of high power microwave and millimetre wave signals, it is noteworthy that higher output optical power level could be obtained at the resonator output as well as higher microwave power level, especially if high linearity photodiode and high power handling optical couplers are used.

Overall, one can see that the FRR selective and repetitive filtering approach features operational superiority over the Fabry-Pérot cavity approach regarding the relative complexity of the Fabry-Pérot cavity setup (dispersion compensation, light injection, coupling, stabilization, etc.) compared to the FRR. Moreover, one must consider the importance of the FRR device if used with the Vernier effect to provide large bandwidth and high repetition rate frequency combs for spectrographs calibration. Still, different noise characterizations are still needed to evaluate the residual noise of such a device, especially if the device is intended to be used for ultrastable and ultra-pure microwaves generation. In fact, the residual noise of the device should be low enough to not affect the performances of the photonic microwave generator. Such noise characterizations need a complex and independent setup to be performed. Moreover, two identical generators must be built and compared to get accurate measurements of the stability, phase noise and linewidth of the generated microwave signal. In our laboratory, we have already demonstrated the generation of a 10 GHz signal by optical frequency division of an ultrastable laser. A fractional frequency stability below 2 x 10$^{-15}$ at one second and a phase noise level below -104 dBc/Hz and -136 dBc/Hz, respectively, at 1 Hz and 10 kHz offset frequencies from the 10 GHz carrier have been achieved[28]. Several studies are currently performed at the laboratory to highly improve the generator performances and to evaluate the contribution of the repetition rate multiplication device if added to the microwave generator scheme.

## Methods

### Double Directional Coupler Add-Drop Fibre Ring Resonator

The double directional coupler add-drop FRR, shown in Figure 10, is fabricated using two low loss fibred 2x2 optical directional couplers ($C_1$ and $C_2$) linked with single-mode optical fibres (SMFs) by one or more splices. Resonance occurs if the total integrated phase shift of the incident light wave around the resonator ring is an integer multiple of 2π radian. The FRR will thus generate a transverse single resonance comb with a $FSR = c/nL$ that lies in the microwave frequency range for a fibre length $L$ of few meters ($n$ is the refractive index and $c \approx 3\times10^8$ m/s is the speed of light in vacuum). The other FRR main characteristics, illustrated in Figure 10, depend on the couplers coupling coefficients ($\kappa_1$ for $C_1$ and $\kappa_2$ for $C_2$) and the overall losses in the resonator architecture (couplers excess loss $\gamma_{1,2}$, splices loss $\alpha_s$ and fibre linear loss $\alpha_f$). Lowering these losses directly improves the FRR characteristics[29, 30].

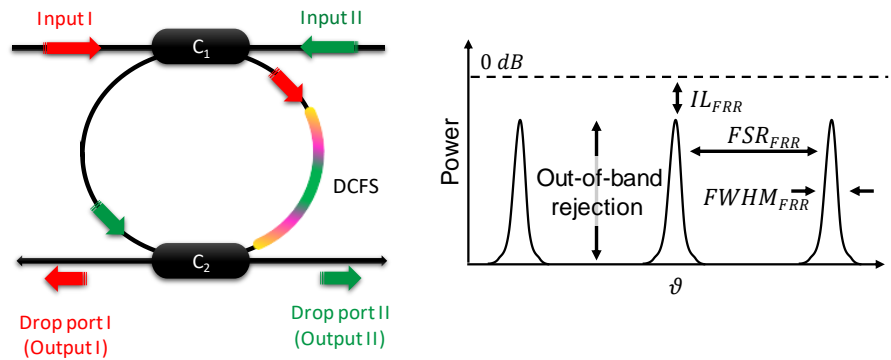

**Figure 10. Double directional coupler fibre ring resonator and the main characteristics of its optical resonance comb.** DCFS: dispersion compensation fibre section; $IL_{FRR}$: insertion loss; $FSR_{FRR}$: free spectral range; $FWHM_{FRR}$: full width at half maximum.

### Dispersion Compensated Fibre Ring Resonator Design and Fabrication

The FRR includes a dispersion compensation fibre section composed of three types of optical fibres: a SMF 28e fibre section, two DCF 3 fibre sections (Vascade LS +) and a DCF 38 fibre section (Vascade S1000). This configuration, illustrated in Figure 10, is used to ensure a zero and flat chromatic dispersion of the FRR over a very large bandwidth. Also, these different fibre sections were accurately calculated, cleaved, and spliced to obtain the desired FRR length. Therefore, the $FSR$ of this hybrid, dispersion compensated FRR is given as follows:

$$FSR_{DCFRR} = \frac{c}{(n_{SMF28e} \cdot L_{SMF28e}) + (n_{DCF3} \cdot L_{DCF3}) + (n_{DCF38} \cdot L_{DCF38})} \qquad (2)$$

Moreover, the FRR dispersion (in ps/nm) is given as follows:

$$D_{DCFRR} = (D_{SMF28e} \cdot L_{SMF28e}) + (D_{DCF3} \cdot L_{DCF3}) + (D_{DCF38} \cdot L_{DCF38}) \qquad (3)$$

If $D_{DCFRR}$ is fixed to 0 ps/nm, from equations (2) and (3) we can get two equations relating the SMF 28 e fibre and DCF 38 fibre sections' lengths to the total DCF3 fibre section length:

$$L_{SMF28e} = \frac{\left(\frac{c}{FSR_{FRR}}\right) - \left(n_{DCF3} - \left(n_{DCF38} \cdot \frac{D_{DCF3}}{D_{DCF38}}\right) \times L_{DCF3}\right)}{n_{SMF28e} - \left(n_{DCF38} \cdot \frac{D_{SMF28e}}{D_{DCF38}}\right)} \quad (4)$$

$$L_{DCF38} = \left(\left(-\frac{D_{SMF28e}}{D_{DCF38}}\right) \times L_{SMF28e}\right) - \left(\left(\frac{D_{DCF3}}{D_{DCF38}}\right) \times L_{DCF3}\right) \quad (5)$$

Considering $n_{DCF3}$ = 1.4695, $D_{DCF3}$ = -0.003 ps/nm.m, $n_{DCF38}$ = 1.4743, $D_{DCF38}$ = -0.0386 ps/nm.m, $n_{SMF28e}$ = 1.4682, $D_{SMF28e}$ = 0.01678 ps/nm.m, at 1550 nm, we can therefore fix $L_{DCF3}$ = 300 mm and calculate $L_{SMF28e}$ = 759.580 mm and $L_{DCF38}$ = 306.954 mm. With these fibre sections' lengths, the dispersion compensated FRR total length is 1.366 m. Accordingly, Figure 11 (a) shows the individual dispersion of the three different optical fibres in function of the wavelength and the dispersion calculated for the compensated FRR. As we can see, a zero and flat dispersion is obtained for the compensated FRR over a large bandwidth.

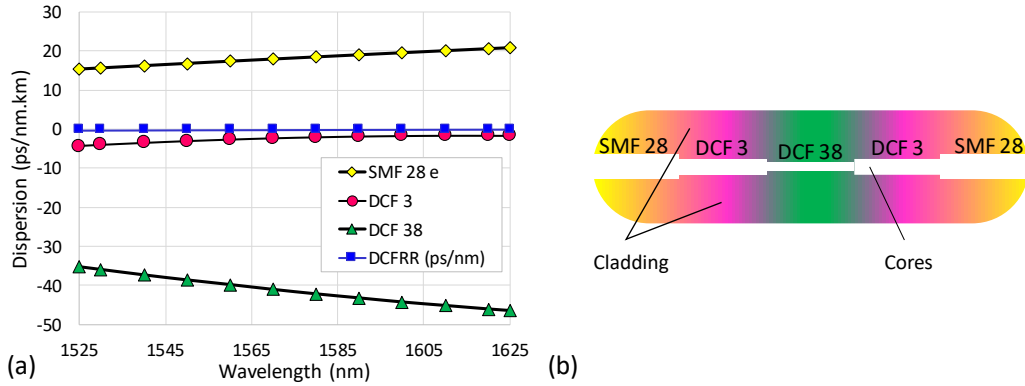

**Figure 11. Dispersion engineering of the FRR.** (a) Evaluation of the dispersion of the dispersion compensated FRR, regarding the dispersion of the different optical fibres used for its fabrication. (b) Illustration (not to scale) showing the profiles of the cores of the different fibres used in the dispersion compensation fibre section included in the FRR. DCFRR: dispersion compensated FRR.

Since the dispersion compensation fibre section introduces additional losses into the FRR, the coupling coefficients of the FRR couplers must be optimized to improve the FRR characteristics[24, 31]. The dispersion compensation fibre section overall loss varies from 0.5 dB up to 1 dB depending on the quality of the splices between the different fibres constituting the dispersion compensation fibre section. This dispersion compensation fibre section loss is due to the difference in the profiles of the cores of the different fibres used [see Figure 11 (b)]. Here, the DCF 3 fibre with a mode field diameter ( $MFD_{DCF3}$ = 8 μm) is used as an intermediate "bridge fibre" to optimize the $MFD$ matching between the SMF 28 e fibre ( $MFD_{SMF28e}$ = 10.4 μm) and the DCF 38 fibre ( $MFD_{DCF38}$ = 5.9 μm), and therefore reduce the section loss and possible light back-reflections. In addition to this technique, and since our Sumitomo T71 fusion splicer does not perform splice tapering, which has been shown to highly improve the quality of the splice between dissimilar fibres[27], we have implemented a technique that ensures a relatively high splice loss reduction factor ( $SLRF$ ) between the dissimilar fibres. The high $SLRF$ was achieved by optimizing the fusion current, fusion time and the number of fusion re-arcs applied after a short first fusion arc. The results of the technique have been confirmed after performing several series of dissimilar fibres splices.

Briefly, after calibrating the fusion arc power and time in the splicer for each splice type (SMF 28e to DCF 3 and DCF 3 to DCF 38), we have used the re-arc function in the fusion splicer which adds an additional short arc to the splice after the first short fusion arc. Therefore, instead of excessively heating both dissimilar fibres, which reacts differently to heat and therefore degrades the $MFD$ matching between fibres, we apply several short re-arcs to the splice to avoid the excessive heating of the fibres and thus improve the $MFD$ matching. The optimal number of re-arcs depends on the splice type. As shown in Figure 12 (a)(b), around 42 re-arcs with a 0.5 second duration are necessary to reduce the SMF 28e to DCF 3 splice loss by 0.08 to 0.11 dB (the results were sensitive to the fibres cleaving angles for this splice type), and around 5 re-arcs with a 0.5 second duration are necessary to reduce the DCF 3 to DCF 38 splice loss by 0.2 dB. Overall, this represents a huge loss reduction regarding the FRR design. We have therefore been able to fabricate such dispersion compensation fibre section with almost 0.5 dB loss (0.1 dB per SMF 28e to DCF 3 splice and 0.15 dB per DCF 3 to DCF 38 splice).

With these parameters, and after the couplers coupling coefficients optimization ($\kappa_1$ = $\kappa_2$ = 10%), the ADS model gives 5.2 dB insertion loss and 20.7 dB out-of-band rejection for the compensated FRR. Thanks to the optimized coupling, an 11 dB optical power ratio gain has been maintained for the compensated FRR, regardless of the loss added by the dispersion compensation fibre section.

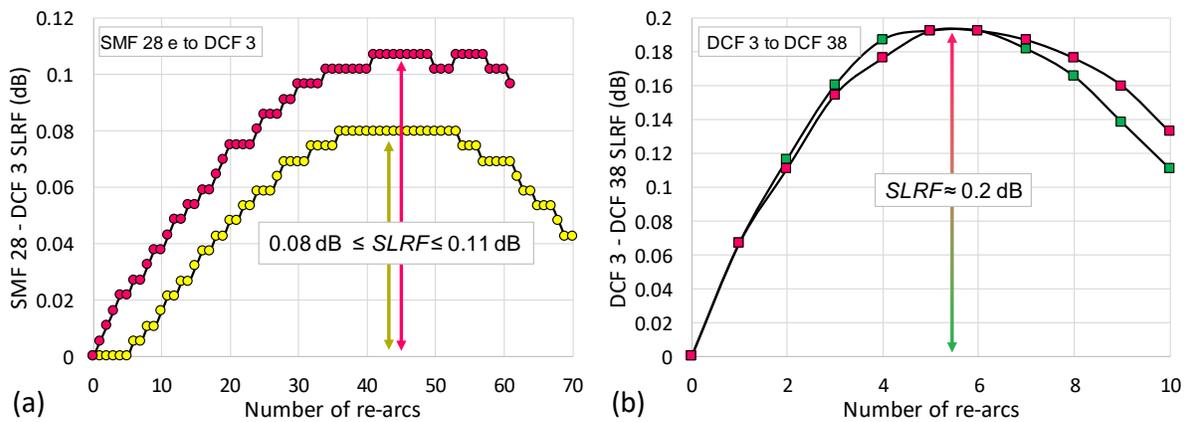

**Figure 12. Dissimilar fibres splice loss reduction.** High splice loss reduction factor ($SLRF$) achieved for (a) SMF 28 e to DCF 3 splice and (b) DCF 3 to DCF 38 splice by optimizing the fusion current, fusion time and the number of fusion re-arcs applied after a short first fusion arc.

Once fabricated, the FRR has been encompassed in an acoustically, mechanically, and thermally well isolated box to reduce these noise contributions. Figure 13 shows a photography of the lab-created compensated FRR.

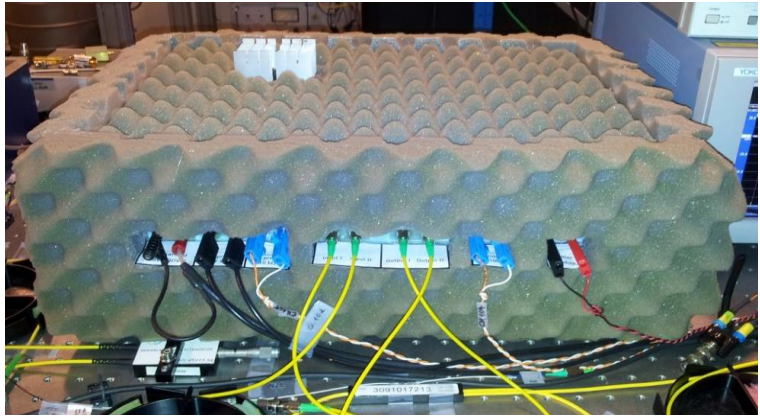

**Figure 13. Dispersion compensated FRR encompassed in an acoustically, mechanically, and thermally isolated box.**

**Optical Resonance Comb Locking**

The ORC locking is achieved using a side-of-fringe locking technique. In this locking scheme, when the ultrastable laser lightwave is at the slope of an optical resonance mode, the optical signal at the FRR second output (see Figure 4) translates the partially-resonant ultrastable laser lightwave's frequency fluctuations into intensity fluctuations. These intensity fluctuations are then fed back to the FRR through a digital proportional–integral controller followed by a Peltier module to control the $FSR$ of the FRR by thermal variations ($\Delta T$). These thermal variations directly affect the fibre refractive index and its length ($\Delta n = 9.2 \times 10^{-6} \times \Delta T$ and $\Delta L = 0.5 \times 10^{-6} \times L \times \Delta T$ in a standard SMF fibre), and therefore the $FSR$ of the resonance comb. As the $FSR$ and the absolute frequencies of resonance modes are linked (the absolute frequency of a given resonance mode $p$ is equal to $p \times FSR$), the stabilization of a given resonance mode to the ultrastable laser ensures (ideally) the stability of the entire ORC in terms of $FSR$ and absolute frequencies of resonant modes (see the illustration in Figure 1). As a result, the ultrastable laser stability is transferred to the entire ORC. However, this stability transfer will be limited by the digital proportional–integral controller performances, the Peltier module responsivity, the ultrastable laser relative intensity noise as well as by the ambient acoustic and mechanical noises.

## Data availability

The datasets generated during and/or analysed during the current study are available from the corresponding author on reasonable request.

**Figures legends**

**Figure 1. Vernier effect obtained by combining the OFC of the femtosecond laser with the ORC of the FRR to selectively preserve or suppress optical modes of the OFC and increase its repetition rate up to a given frequency of interest**. The ultrastable laser (USL) is used to lock the ORC.

**Figure 2. Vernier solutions for different values of** $N_2$. (a) Optical power ratio $\beta$ between the OFC desirable modes, spaced by 10 GHz, and the undesirable modes in-between, calculated for different Vernier solutions over a 10 GHz optical bandwidth. (b) Most evident patterns for different Vernier solutions simulated over a 10 GHz optical bandwidth, while considering an input OFC having a 0 dBm optical power per mode. The mode number scale refers to the OFC modes included in the 10 GHz optical bandwidth, where modes 0 and 40 are the desirable optical modes spaced by 10 GHz and the modes from 1 to 39 are the undesirable optical modes in-between.

**Figure 3. Simulated Vernier patterns obtained for the 10 GHz Vernier solution, while considering an input OFC having a 0 dBm optical power per mode**. (a) Some of the patterns of the 10 GHz Vernier solution, where a symmetry in the patterns repeatability is noticeable, whether the $N_2$ value is higher or lower than $N_1$ value (40) by the same integer number (the solution where $N_2$ = 40 is exposed in the back for more visibility of the symmetry in the patterns repeatability). (b) The Vernier pattern obtained when $N_2$ = 67, compared to the case where no FRR is used.

**Figure 4. Experimental setup used to characterize the OFC's $f_{rep}$ multiplication by Vernier effect using the FRR.** OFC: femtosecond laser optical frequency comb; DCFS: dispersion compensation fibre section; VOA: variable optical attenuator; PC: polarization controller; FP: Fabry-Pérot cavity; USL: ultrastable laser; OC: optical fibre coupler; PM: power meter; OSA: optical spectrum analyser; PD: photodiode; OSC: ultrafast oscilloscope; ESA: electrical spectrum analyser; PI: digital proportional–integral controller.

**Figure 5. Optical domain characterization.** (a) Optical spectra recorded with and without the compensated FRR, at 3 dBm incident average optical power, confirming a total conservation of the OFC linewidth at the compensated FRR output. (b) Zoom into the recorded optical spectra showing the 10 GHz Vernier pattern formation. OSA: RBW = 2.5 GHz.

**Figure 6. Microwave domain characterization.** Microwave spectra recorded (a) without and (b) with the compensated FRR, at 3 dBm incident average optical power, confirming the chosen 10 GHz Vernier pattern formation. ESA: RBW = 100 kHz and VBW = 100 kHz. DCFRR: dispersion compensated FRR.

**Figure 7. Time domain characterization.** (a) Electrical waveforms recorded with the ultrafast oscilloscope, with and without the compensated FRR, at 3 dBm incident average optical power. (b) A zoom into the pulses at the compensated FRR output reveals the presence of the 10 GHz frequency and its harmonics (0.1 ns period). Ultrafast oscilloscope sampling rate = 80GS/s.

**Figure 8. Microwave power characterization.** Microwave power measurements performed at 10 GHz versus the incident average optical power, obtained with and without the compensated FRR. The case of an ideal photodetection is added for comparison.

**Figure 9. Simulated Vernier patterns obtained when $N_2$ = 39 with a compensated FRR and a zero-dispersion FRR, compared to the case where no FRR is used, considering an input OFC having a 0 dBm optical power per mode**. The mode number scale refers to the OFC modes included in the 10 GHz optical bandwidth, where modes 0 and 40 are the desirable optical modes spaced by 10 GHz and the modes from 1 to 39 are the undesirable optical modes in-between. DCFRR: dispersion compensated FRR; ZDFRR: zero-dispersion FRR.

**Figure 10. Double directional coupler fibre ring resonator and the main characteristics of its optical resonance comb.** DCFS: dispersion compensation fibre section; $IL_{FRR}$: insertion loss; $FSR_{FRR}$: free spectral range; $FWHM_{FRR}$: full width at half maximum.

**Figure 11. Dispersion engineering of the FRR.** (a) Evaluation of the dispersion of the dispersion compensated FRR, regarding the dispersion of the different optical fibres used for its fabrication. (b) Illustration (not to scale) showing the profiles of the cores of the different fibres used in the dispersion compensation fibre section included in the FRR. DCFRR: dispersion compensated FRR.

**Figure 12. Dissimilar fibres splice loss reduction.** High splice loss reduction factor ($SLRF$) achieved for (a) SMF 28 e to DCF 3 splice and (b) DCF 3 to DCF 38 splice by optimizing the fusion current, fusion time and the number of fusion re-arcs applied after a short first fusion arc.

**Figure 13. Dispersion compensated FRR encompassed in an acoustically, mechanically, and thermally isolated box.**

## Acknowledgements (not compulsory)

The work has been performed in the frame of the ANR project "Projets d'Investissements d'Avenir (PIA) Equipex Oscillator-Imp". The authors would like to thank the Council of the Région de Franche-Comté for its support to the PIA and the Labex FIRST-TF for funding.

## Author contributions statement

K.S. performed the simulations, designed, and fabricated the device, K.S., J.M., B.M., B.D., A.B., A.D., C.L. and Y.K. conducted the experiments, K.S. wrote the manuscript, K.S., J.M., B.M., B.D., A.B., A.D., C.L. and Y.K. reviewed the manuscript and agreed on the interpretations.

## Additional information

**Competing financial interests:** The authors declare no competing financial interests.